\documentclass[10pt]{article}

\usepackage{wrapfig}           
\usepackage{graphicx}          
\usepackage{dcolumn}           
\usepackage{bm}                
\usepackage{float}             

\title{Physico-mathematical foundations of relativistic cosmology}

\author{Domingos Soares \\ {\it Physics Department}\\
{\it Universidade Federal de Minas Gerais} \\   
{\it Belo Horizonte, Brazil} } 

\date{September 09, 2013}

\begin{document}

\maketitle


\begin{abstract}
I briefly present the foundations of relativistic cosmology, which are, 
General Relativity Theory and the Cosmological Principle. I discuss some 
relativistic models, namely, ``Einstein static universe" and ``Friedmann 
universes". The classical bibliographic references for 
the relevant tensorial demonstrations are indicated whenever necessary, although 
the calculations themselves are not shown.
\end{abstract}


\section{Introduction}

The {\it Cosmological Principle} (CP) --- the homogeneity and isotropy of the 
universe --- and the {\it General Relativity Theory} (GRT)  are the physico-mathematical 
foundations of relativistic cosmology. Summarizing all that will be presented in the following sections, 
one can simply state that the symmetries introduced by the CP reduce Einstein's full field equations 
of GRT to two straightforward differential equations in the scale --- or expansion --- factor of the 
universe \cite[p. 260]{harw}. The most popular relativistic cosmological models are built from 
these two equations.    

Einstein's field equations of GRT represent a mathematical description of a geometrical 
entity, the space-time, defined by three spatial coordinates and a temporal one. Such a 
four-dimensional entity is established by the existing matter and energy contents. On the left-hand   
side of the equations one has the geometrical description of space-time and on the right-hand  
side, the energy and momentum contents. Putting it in another way, GRT is Einstein's theory of 
gravitation. It can be  understood, in a simplified way, by stating that {\it ``Spacetime grips mass, 
telling it how to move; and mass grips spacetime, telling it how to curve"} \cite[p. 275]{tawh}. 
This is obviously an incomplete statement since not only matter curves space-time but also all 
kinds of energy \cite[p. 229]{harr}.  

Einstein uses the tensorial formalism to express his field equations, therefore, GRT is a tensorial 
theory. Incidentally, the German mathematician Georg Friedrich Bernhard Riemann (1826-1866) 
was one of the main responsible for the development of {\it tensor calculus}, playing an 
extraordinary role in the formulation of GRT.  But, what is a {\it tensor}? 
A tensor is a mathematical entity that has in each point of space $n^m$ components, where $n$ 
is the {\it number of dimensions of space} and $m$ is the {\it order of the tensor}. Hence, one 
can say that the {\it scalar} is a tensor of order 0 --- therefore, it has 1 component --- and that the 
{\it vector} is a tensor of order 1 --- it has n components \cite[p. 200]{harr}. Tensors used in GRT 
are tensors of order m=0,1 and 2 and ``space" is the space-time of n=4 dimensions (three 
spatial coordinates and one temporal coordinate). Therefore, the second-order tensors of GRT 
have, in principle, $4^2=16$ components. I say ``in principle" because real physical problems 
impose symmetry constraints which reduce to 10 the really necessary components. The first-order 
tensors are the GRT vectors, called quadrivectors and have 4 components. 

I shall make a simplified presentation of Einstein's field equations which consist of 10 nonlinear 
differential equations  --- a system of equations. This system is tremendously simplified when 
additional symmetry constraints imposed by the CP (see section 2 of  \cite{hwan}) are considered 
and reduce itself to 2 equations --- only 2 components of GRT's tensors are necessary to the 
complete formulation of the modern relativistic cosmology of the ``Hot Big-Bang" theory.  

Einstein's field equations of GRT can be qualitatively expressed \cite[p. 229]{harr} as:

\bigskip
\begin{equation}
\label{eq:ee1}
\mbox{space-time curvature = constant} \times \mbox{matter-energy.} 
\end{equation}
\bigskip

Space-time curvature is mathematically given by the {\it Einstein tensor} $G_{\mu\nu}$:

\bigskip
\begin{equation}
\label{eq:ee2}
{G}_{\mu\nu} = R_{\mu\nu} - \frac{1}{2} g_{\mu\nu} R,  
\end{equation}
\bigskip

\noindent with indices $\mu$ and $\nu$ assuming the values of 0,1,2 and 3. The tensor 
$R_{\mu\nu}$ is called {\it Ricci tensor}, formed from the {\it Riemann curvature tensor}, 
which is a tensor of order 4, being the most general way of describing the curvature of any 
n-dimensional space. In the case of GRT, the space-time of 4 dimensions implies in the existence 
of $4^4=256$ components. The Ricci tensor, of order 2, is the reduced form of the Riemann tensor 
to be used in Einstein's equations. The reduced form is obtained through the application of symmetry 
relations that eliminate the redundant terms in the Riemann tensor. The tensor $g_{\mu\nu}$ is the 
space-time {\it metric tensor} and play, in Einstein's equations, the role of the {\it field} 
\cite[p. 179]{rind}. And that is why one says ``Einstein's {\it field} equations". We do not  
speak, in GRT, of ``action at a distance"; a test particle does not directly ``feel" the sources of 
matter and energy, rather it feels the field, i.e., the metric --- the geometry --- that such sources 
generate in its neighborhood. The metric field transmits the perturbations in the geometry 
(gravitational waves) at the speed of light, a similar situation to what happens in 
electromagnetism \cite[p. 179]{rind}. The metric field is the analogue of the gravitational field 
in Newtonian theory. Finally, the term R, in Eq. 2, is the {\it scalar curvature}, a scalar associated 
to the Ricci tensor and to the metric. In tensorial terms, the scalar curvature is equal to the 
{\it trace of the Ricci tensor with respect to the metric tensor}. The scalar curvature is also 
called the {\it Ricci scalar} \cite[p. 219]{rind}.   

The part of matter and energy in Einstein's equations is given by the {\it energy-momentum tensor} 
$T_{\mu\nu}$. The full Einstein's field equations have, thus, the following compact form:    

\bigskip
\begin{equation}
\label{eq:ee3}
{G}_{\mu\nu} = -\kappa T_{\mu\nu},  
\end{equation}
\bigskip

\noindent where $\kappa = 8\pi G/c^4$ is the {\it Einstein gravitational constant}, G is the 
universal gravitational constant and c is the speed of light in vacuum. Finally, inserting  Eq. \ref{eq:ee2} 
into \ref{eq:ee3}, one has the explicit form of the 16 equations of Einstein: 

\bigskip
\begin{equation}
\label{eq:ee4}
R_{\mu\nu} - \frac{1}{2} g_{\mu\nu} R = -\frac{8\pi {\rm G}}{c^4} T_{\mu\nu}. 
\end{equation}
\bigskip

And in matrix format, still with the 16 components, one has:

\bigskip
\begin{eqnarray*}
\left(\begin{array}{cccc}
R_{00}&R_{01}&R_{02}&R_{03}\\
R_{10}&R_{11}&R_{12}&R_{13}\\
R_{20}&R_{21}&R_{22}&R_{23}\\
R_{30}&R_{31}&R_{32}&R_{33}
\end{array}\right)
-\frac{1}{2}
\left(\begin{array}{cccc}
g_{00}&g_{01}&g_{02}&g_{03}\\
g_{10}&g_{11}&g_{12}&g_{13}\\
g_{20}&g_{21}&g_{22}&g_{23}\\
g_{30}&g_{31}&g_{32}&g_{33}
\end{array}\right)
R = 
\end{eqnarray*}

\bigskip

\begin{equation}
\label{eq:ee5}
= -\frac{8\pi {\rm G}}{c^4}
\left(\begin{array}{cccc}
T_{00}&T_{01}&T_{02}&T_{03}\\
T_{10}&T_{11}&T_{12}&T_{13}\\
T_{20}&T_{21}&T_{22}&T_{23}\\
T_{30}&T_{31}&T_{32}&T_{33}
\end{array}\right). 
\end{equation}
\bigskip

Before discussing Einstein's field equations, shown above, I describe, in the following section,  
the term that appears on the right-hand side of Eqs. \ref{eq:ee3}, \ref{eq:ee4}  
and \ref{eq:ee5}, namely, the energy-momentum tensor $T_{\mu\nu}$.  I discuss in section 3 the  
GRT equations, first, the vacuum equations, i.e., in the absence of sources of matter and energy, 
and then the full equations, which are the relevant equations for cosmology. Next, in section 4, 
I pass to the cosmological applications of the full equations.  Here, I add a novelty, the 
inclusion of the so-called {\it cosmological constant} $\Lambda$ in Eq. \ref{eq:ee4}, which increases 
its generality. Einstein's static universe --- where the cosmological constant explicitly appears --- 
is discussed in section 4.1 and the Friedmann universes, in which $\Lambda$=0, are shown in 
section 4.2. In section 5, I present some final remarks.

It is necessary, at this point, to make an important warning. It is not possible a complete and 
satisfactory apprehension  of GRT without the knowledge --- even a rudimentary one --- of the 
techniques of tensor calculus. But it is possible, nevertheless, to have a general idea of the 
theoretical picture, even without going deeply into the tensorial demonstrations. That's exactly what 
I intend, in this presentation of one of the most known applications of GRT, namely, modern 
cosmology. When they are necessary, I always  give the appropriate references for the reader 
interested in the tensorial mathematical formalism.    

\section{Energy-momentum tensor}

This is the tensor --- the right-hand side of  Eq. \ref{eq:ee4} --- that describes the energetic activity 
in space. The energy-momentum tensor quantitatively gives the {\it densities} and the  
{\it fluxes} of energy and momentum generated by the sources present in space and which will 
determine the geometry of space-time --- the left-hand side of  Eq. \ref{eq:ee4}.   

The components of the energy-momentum tensor \cite[p. 137]{mtw} are the following:
\begin{description}
\item{$T_{00}$ =} density of matter and energy.
\item{$T_{0\nu}$ =} flux of energy (i.e., energy per unity of  
area, per unity of time) in the $\nu$ direction; $\nu\ne 0$.
\item{$T_{\mu 0}$ =} density of the $\mu$ component  of momentum; $\mu\ne 0$.
\item{$T_{\mu\nu}$ =} flux of the $\mu$ component  of momentum in the $\nu$ direction  
(i.e., shear stress). Note that ``flux of momentum" is the same as ``force per area"; 
$\mu,\nu\ne 0$.
\item{$T_{\mu\mu}$ =} flux of the $\mu$ component  of momentum in the $\mu$ direction  
(i.e., force over the perpendicular area, that is, pressure, which differs from shear stress precisely 
for taking into account the component of the force perpendicular to the surface upon which it 
acts); $\mu\ne 0$.  
\end{description}
The tensor of energy-momentum is symmetric, that is, $T_{\mu\nu}$ =  
$T_{\nu\mu}$. Misner, Thorne and Wheeler \cite[p. 141]{mtw} show this using a physical argument. 
They consider the shear stresses upon a small cube of side L and mass-energy $T_{00}L^3$ and 
go on to show that it would have infinite angular acceleration if the tensor was not symmetric. 
 
Being symmetric, the energy-momentum tensor has a maximum of 10 different components, 
instead of the 16 of any $4\times4$ tensor. As we shall see in what follows, the tensors 
related to the geometry of space-time, in the left-hand side of  Eq. \ref{eq:ee4}, are also 
symmetric. 

We are ready now to discuss, in more details, Einstein's equations of GRT.
                 
\section{Einstein's field equations}

The enormous success of Newton's gravitation in classical phenomena --- weak gravitational 
fields and velocities much smaller than the speed of light --- make it almost {\it mandatory} 
that any new theory of gravitation reduce itself, in those limits, to the Newtonian law of the 
inverse square.  In other words, in the so-called ``classical limit", in the absence of gravitational 
sources, the GRT must reduce to the {\it Laplace equation} for the Newtonian gravitational potential 
$\Phi$, $\nabla^2\Phi=0$, and to the {\it Poisson equation},  $\nabla^2\Phi=4\pi G\rho$, whenever 
there are sources, represented by the density of matter $\rho$.  

This was the course followed by Einstein, which I pass to discuss, first, with the field equations 
in the absence of sources and, then, with the full equations, that have as special case the first 
ones.

Besides the reduction to the Newtonian limits, Einstein used also, for the postulation of the field 
equations, the criteria of {\it simplicity} and of {\it physical intuition}. Einstein asks himself: {\it 
What is the most simple form of the space-time metric, in the absence of sources, that produces, in  
the classical limit, the Laplace equation for the Newtonian gravitational potential?} {\it And if there 
are sources, how to get the most simple form of the metric tensor and, at the same time, 
the classical reduction to the Poisson equation?} And yet more, in both cases of the general theory, 
he should obtain the conservation of energy and of momentum. 

After the fulfillment  of such criteria and of the reduction to the classical limits, the validity of the 
formulated equations must, of course, be verified by experiments. As we shall see, such a verification 
occurred in an extremely satisfactory way for the vacuum equations, but, apparently, has not yet 
occurred for the full equations.   
                  
\subsection{Vacuum equations}

GRT's vacuum equations are in grand manner --- and to a certain point, paradoxically, for not being 
the full field equations --- the great responsible for the extraordinary prestige enjoyed by GRT. That 
is what we shall see in what follows. 

The vacuum equations are those valid for the metric field in vacuum, as for example, for the field 
around the Sun, where the density of matter  $\rho=0$. By investigating the symmetries of the 
Ricci tensor  $R_{\mu\nu}$ in the classical limit of the metric tensor $g_{\mu\nu}$ \cite[p. 222]{rind}, 
Einstein postulates the following form for the vacuum field equations:

\bigskip
\begin{equation}
\label{eq:ee6}
R_{\mu\nu} = 0. 
\end{equation}
\bigskip

Einstein put forward the vacuum equations of GRT in 1915, and in 1916 the German astronomer and 
physicist Karl Schwarzschild (1873-1916) derived the first and the most important exact solution 
of the vacuum field equations, known as the {\it Schwarzschild metric} \cite[p. 228]{rind}. Such a 
solution has been applied with great success, for example, to planetary motion, giving the 
right explanation for the phenomenon of the precession of Mercury's orbit --- which was not 
obtained with Newtonian gravitation --- and predicting new phenomena, amongst them, the 
deflection of a light ray traveling next to a large matter concentration  \cite[p. 223]{rind}. These, 
and other tests, were realized with great experimental success and are responsible for the almost 
unanimous acceptance of GRT by the scientific community. Schwarzschild's metric is also responsible 
for the current and controversial discussion of  phenomena such as gravitational radiation and 
{\it black holes}.   

These are not, however, the field equations that will lead to the modern models  of relativistic 
cosmology. The appropriate field equations  need the presence of sources of matter and of radiation 
to be properly applied to the universe. Such equations, called {\it full}, had not yet definitive 
experimental confirmation (see discussion in  \cite{so09a}) and are presented below.    
 
\subsection{Full field equations}

The first attempt for the field equations in the presence of sources would be obviously a modification 
of Eq. \ref{eq:ee6}, i.e., 
$R_{\mu\nu} = \mbox{constante} \times T_{\mu\nu}$, which does not work because the 
divergent of  $R_{\mu\nu}$ is not null, implying that there is no conservation of energy 
and momentum. The second choice is simply substituting $R_{\mu\nu}$ 
by the Einstein tensor $G_{\mu\nu}$ (Eq. \ref{eq:ee2}) which does have 
null divergent \cite[p. 299]{rind}. The full field equations then take the form of Eq. \ref{eq:ee4} and, 
in matrix format, of Eq. \ref{eq:ee5}. But there are additional simplifications. We have seen that 
the energy-momentum tensor is symmetric.  And due to space-time symmetries --- such as, the least 
distance from A to B is the same as from B to A, and, the distance around a circle is the same 
in the clockwise and anticlockwise directions ---  the same is true for the Ricci and metric tensors 
\cite[p. 240]{harr}, and they have, just like it is the case for $T_{\mu\nu}$, at most 10 different 
components in each event of space-time.  

Einstein's full field equations are hence written as 
$G_{\mu\nu} \equiv R_{\mu\nu} - 1/2g_{\mu\nu}R = -\kappa T_{\mu\nu}$, where $\kappa$ 
is Einstein's gravitational constant, defined in section 1. In matrix format, we have:

\bigskip
\begin{eqnarray*}
\left(\begin{array}{cccc}
R_{00}&R_{01}&R_{02}&R_{03}\\
=&R_{11}&R_{12}&R_{13}\\
=&=&R_{22}&R_{23}\\
=&=&=&R_{33}
\end{array}\right)
-\frac{1}{2}
\left(\begin{array}{cccc}
g_{00}&g_{01}&g_{02}&g_{03}\\
=&g_{11}&g_{12}&g_{13}\\
=&=&g_{22}&g_{23}\\
=&=&=&g_{33}
\end{array}\right)
R = 
\end{eqnarray*}

\bigskip

\begin{equation}
\label{eq:ee7}
= -\frac{8\pi {\rm G}}{c^4}
\left(\begin{array}{cccc}
T_{00}&T_{01}&T_{02}&T_{03}\\
=&T_{11}&T_{12}&T_{13}\\
=&=&T_{22}&T_{23}\\
=&=&=&T_{33}
\end{array}\right).
\end{equation}
\bigskip

It is interesting to note that the full field equation is reduced to the vacuum equation when 
$T_{\mu\nu}=0$. This happens in the following way. Einstein's full field equation can also be 
written in the form $R_{\mu\nu} = -\kappa(T_{\mu\nu}-1/2g_{\mu\nu}T)$  \cite[p. 299]{rind}. 
It becomes apparent, therefore, that for $T_{\mu\nu}=0$ one has $R_{\mu\nu}=0$, 
that is, Eq. \ref{eq:ee6}.
                  
\section{Cosmological models}

The cosmological models are built up by means of Einstein's full field equations.  One of the most 
tedious and laborious tasks in GRT is the calculation of the Ricci tensor, the scalar curvature and  
finally the Einstein tensor $G_{\mu\nu}$, a calculation made for a given metric 
tensor $g_{\mu\nu}$.   

Rindler \cite[p. 418]{rind} shows how these calculations must be done for a generic diagonal 
metric given by: 

\bigskip
\begin{equation}
\label{eq:ee8}
(ds)^2 = A(dx_0)^2 + B(dx_1)^2 + C(dx_2)^2 + D(dx_3)^2, 
\end{equation}
\bigskip

\noindent where A, B, C and D are arbitrary functions of all space-time coordinates. 

Let us now obtain the equations of relativistic cosmology. The CP, that is, the reduction of the 
real universe to a homogeneous and isotropic idealization, implies in a space-time with the 
Robertson-Walker metric \cite[p. 367]{rind}, which, in spatial spherical coordinates, is given by: 

\bigskip
\begin{equation}
\label{eq:ee9}
(ds)^2 = (cdt)^2 - S^2(t)\left[\left(dr\over{\sqrt{1-kr^2}}\right)^2 +  
(rd\theta)^2 + (r{\rm sen}\theta d\phi)^2\right]. 
\end{equation}
\bigskip

S(t) is the {\it scale factor} of the universe and k is the {\it spatial curvature constant}. 
This equation is used, instead of Eq. \ref{eq:ee8}, for the calculation of the Einstein tensor 
applied to cosmology, i.e., to the idealized universe of the CP. 

In order to increase the generality of the field equations, one adds to the scalar curvature a 
constant, the so-called {\it cosmological constant} $\Lambda$, that will be essential for the 
discussion of Einstein's static universe, in the following section. Such a constant is called 
{\it cosmological} because it is only relevant in the context of cosmology, i.e., for the 
structure and evolution of the universe. The full field equation can be written then as 

\bigskip
\begin{equation}
\label{eq:ee10}
R_{\mu\nu} - (\frac{1}{2}R  - \Lambda)g_{\mu\nu} = -\kappa T_{\mu\nu}. 
\end{equation}
\bigskip

The cosmological constant does not change in anything the formal validity of the field equations, 
and can be positive, negative or null. In the last case, of course, one recovers the usual 
formulation of the field equations (Eq. \ref{eq:ee4}). According to Rindler \cite[p. 303]{rind} 
{\it ``The  $\Lambda$ term seems to be here to stay; it belongs to the field equations much 
as an additive constant belongs to an indefinite integral."} Like the scalar curvature R, $\Lambda$ 
has dimensions of length$^{-2}$.

Just like the full field equation without $\Lambda$, the full field equation with $\Lambda$ is reduced 
to the vacuum equation when $T_{\mu\nu}=0$. As before, Einstein's full field equation can also be 
written in the form $R_{\mu\nu}=-\kappa(T_{\mu\nu}-1/2g_{\mu\nu}T)+
g_{\mu\nu}\Lambda$. For $T_{\mu\nu}=0$ one has \cite[p. 303]{rind}: 

\bigskip
\begin{equation}
\label{eq:ee11}
R_{\mu\nu} = g_{\mu\nu}\Lambda. 
\end{equation}
\bigskip

This equation, for the vacuum, that substitutes Eq. \ref{eq:ee6}, without $\Lambda$, is 
only important for eventual cosmological studies. It is totally irrelevant, for example, in solar 
system studies.  In such a case, Eq. \ref{eq:ee6}, and its solution, Schwarzschild's metric, 
is perfectly satisfactory, even if there exists a cosmological constant. 

In order to get the relativistic cosmological equations one has to make the fundamental assumption 
of the CP: all matter --- including a possible ``dark matter" --- of the universe will be, so to speak, 
{\it pulverized} and redistributed in an uniform way throughout the universe. One has in such a way 
the physical requirements of the CP, i.e., the homogeneity and the isotropy of matter distribution. 
The energy-momentum tensor of these sources, namely, matter and radiation with the physical 
characteristics of uniformity, is reduced to the diagonal elements \cite[p. 392]{rind} 
\cite[p. 140]{mtw}:       

\bigskip
\begin{equation}
\label{eq:ee12}
T_{\mu\nu} = \mbox{diag}(\rho c^2, -p, -p, -p), 
\end{equation}
\bigskip

\noindent where p is the {\it isotropic} pressure and $\rho$ is the {\it homogeneous} density 
of the fluid. A fluid like this is called a {\it perfect fluid}. The negative sign that appears in p implies, 
in the field equation, that a {\it positive pressure} has an {\it attractive gravitational effect} 
\cite[p. 156]{rind} \cite[p. 172]{hbn}.

One must notice also that p represents the pressure of radiation and matter, and, in the same 
way, $\rho$ must be split up into one part for radiation and one part for matter. Radiation pressure 
will only be significant in the initial stages of expanding models --- a photon gas at high temperature 
with $p=1/3\rho c^2$.  Matter, which only appears in later stages, has negligible pressure. A perfect 
fluid of matter with zero pressure is often, technically, called {\it dust}. Such a dust stays at rest 
in the spatial substratum, because any random motion would result in the existence of pressure. 
Global motions of expansion or contraction are not excluded, though.  

The main work, in order to obtain the cosmology equations, is to apply the left-hand side of the 
field equations given by Eq. \ref{eq:ee10} --- the Einstein tensor --- to the Robertson-Walker metric 
given by Eq. \ref{eq:ee9}. It has been already said above that Rindler \cite[p. 418]{rind} shows 
the detailed calculations needed to obtain each element of Einstein's tensor. Schwarzschild's metric, 
being the metric of a homogeneous and isotropic universe, with its innumerable symmetries, 
implies that  the Einstein tensor, with the cosmological constant, 
 $G_{\mu\nu}\equiv R_{\mu\nu}-(1/2R-\Lambda)g_{\mu\nu}$, will only have the diagonal 
elements, precisely like the energy-momentum tensor: 
 
\bigskip
\begin{equation}
\label{eq:ee13}
G_{\mu\nu} = \mbox{diag}(G_{00}, G_{11}, G_{22}, G_{33}). 
\end{equation}
\bigskip

Rindler \cite[p. 392]{rind} and Misner, Thorne and Wheeler \cite[p. 728]{mtw} give the results of 
the calculations for $G_{\mu\nu}$:

\bigskip
\begin{equation}
\label{eq:ee14}
G_{00}=-{3\dot{S}^2\over S^2c^2}-{3k\over S^2} + \Lambda, 
\end{equation}
\bigskip

\begin{equation}
\label{eq:ee15}
G_{11}=G_{22}=G_{33}=-{2\ddot{S}\over Sc^2}-{\dot{S}^2\over S^2c^2}
-{k\over S^2} + \Lambda, 
\end{equation}
\bigskip

\noindent where S(t) is the scale factor of the universe (cf. Eq. \ref{eq:ee9}).

With these values for $G_{\mu\nu}$ and $T_{\mu\nu}$, the full Einstein equations, 
$G_{\mu\nu}=-8\pi G/c^4T_{\mu\nu}$  (Eq. \ref{eq:ee10}), reduce themselves to only 
two non-linear differential equations for the scale factor S(t):  

\bigskip
\begin{equation}
\label{eq:ee16}
{\dot{S}^2\over S^2c^2}+{k\over S^2} - {\Lambda\over 3}=
{8\pi G\rho\over 3c^2}, 
\end{equation}
\bigskip

\begin{equation}
\label{eq:ee17}
{2\ddot{S}\over Sc^2}+{\dot{S}^2\over S^2c^2}+{k\over S^2}-\Lambda =
-{8\pi Gp\over c^4}. 
\end{equation}
\bigskip

Eqs. \ref{eq:ee16} and \ref{eq:ee17} are the basic equations for the formulation of the majority 
of relativistic cosmological models, with or without the cosmological constant. Moreover, the 
simultaneous solution of this system of equations leads to the equation of mass and energy 
conservation of the cosmic fluid \cite[p. 393]{rind}.

Next, I show a cosmological model with $\Lambda$ and a family of models without $\Lambda$. 
          
\subsection{Einstein's static universe}

Soon after the final presentation of the GRT, in 1915, Albert Einstein (1879-1955) inaugurated 
the study of relativistic cosmology. He published, in 1917, an article with the suggestive title 
{\it ``Cosmological considerations on the General Theory of Relativity"}. He uses  Eq. \ref{eq:ee16} 
with positive $\Lambda$ in order to get a repulsive cosmic effect and, hence, exactly counterbalance 
the attractive effect of matter and radiation of the universe. Thus, he obtains a static universe, which 
was in accordance with the current ideas of the epoch. This model was of enormous importance 
in the history of the science of cosmology because it was a source of scientific inspiration for 
many investigators. The model had, however, an undesirable feature: it was unstable under small 
perturbations on the state of static equilibrium. Einstein's model is discussed in detail by Soares 
\cite{so12}, including its instability. Eq. 2 of Soares  \cite[p. 1302-2]{so12} is the same 
Eq. \ref{eq:ee16} determined here.
          
\subsection{Friedmann's universes}

The Russian physicist, meteorologist and cosmologist Aleksandr Aleksandrovich Friedmann 
(1888-1925) was responsible for the next great contribution to relativistic cosmology. In 1922, 
he published an article, in a prestigious German scientific journal, with the title {\it ``On the 
curvature of space"}, where he solves Einstein's full field equations, with the hypotheses 
of homogeneity and isotropy of the universe --- which would turn out to be later known as the 
``Cosmological Principle" --- and obtains a model of positive spatial curvature (spherical space) 
with expanding and contracting phases. Subsequently, it was recognized that this was only one 
of the possibilities of dynamical universes --- an oscillating closed universe --- amongst others,  
namely, the open universes; these are one of negative curvature, hyperbolic space, and another 
of null curvature, flat or Euclidean space. 

Friedmann's work was only recognized by the scientific community a long time after its 
publication. In his homage, the models resulting from  Eq. \ref{eq:ee16}, without $\Lambda$, 
were called {\it Friedmann's models} or {\it universes}.  A detailed discussion, albeit at the  
elementary level, on Friedmann's models is presented in  \cite{viso}.   
                  
\section{Final remarks}

Einstein's static model was the first relativistic cosmological model and, also, the first to 
use the cosmological constant. Modern relativistic cosmological models also adopt the 
cosmological constant, and by its means are able to get consistency between the theoretical 
age of the universe and the limits imposed by stellar evolution. In other words, the 
cosmological constant is able to solve the so-called {\it age of the universe dilemma} 
(more details in \cite{so09b}).   

It is worthwhile pointing out that Friedmann (section 4.2) originally obtained only the closed 
model, by means of Einstein's full field equations and of the CP, which was a cosmological 
concept introduced by him. The three modern {\it Friedmann models} appeared with the 
generalization introduced by the Robertson-Walker metric  (Eq. \ref{eq:ee9}), which predicts 
still other global spatial topologies, specified by the spatial curvature constant k, besides the 
known open models, hyperbolic, with k=--1, and flat, with k=0 \cite[p. 367]{rind}.

As anticipated in \cite{so09a}, we saw, in section 3.1, that the great and decisive tests of GRT are 
done for one solution of Einstein's field equations for the {\it vacuum}, that is, in the absence of 
sources of energy and momentum. Such a solution has innumerable practical applications and is 
given by Schwarzschild's metric. The most known solutions of the {\it full} field equations are 
precisely the relativistic cosmological models, valid for a homogeneous and isotropic fluid, and 
they fail when confronted with observations. The relativistic models only survive when the existence 
of unobservable physical entities like dark matter and dark energy are postulated  (cf. \cite{so09a}).

\end{document}